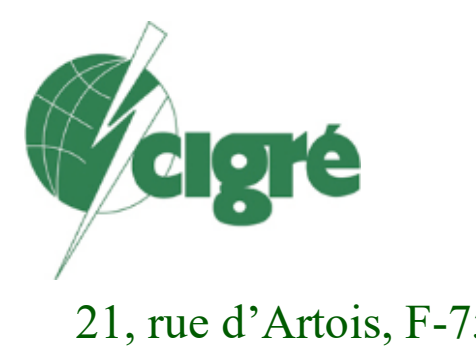

# Grid-Forming E-STATCOMs for Stable Integration of Large-Scale Data Centers: Modeling and Control

**P. R. BANA[†], N. ZAKKIA[††], J. P. HASLER[†], and C. DANIELSSON[†]**
**Hitachi Energy**
**Sweden[†], USA[††]**

## SUMMARY

The rapid expansion of large-scale AI data centers (AIDC) is introducing new stability challenges, particularly in weak or low-inertia networks characterized by fast, step-like demand variations and strict requirements on voltage and dynamic performance. This paper investigates the use of grid-forming (GFM) Enhanced STATCOMs (E-STATCOMs) to support reliable integration of such facilities. A power-admittance-based linear modelling framework is developed to capture system interactions and is validated through detailed EMT simulations. The results demonstrate that E-STATCOMs provide fast, well-damped responses to abrupt load changes while effectively mitigating low-frequency oscillations and interactions with network resonances. By enabling tunable dynamic behavior via a load balancer, virtual impedance, and coordinated active-reactive power support, the proposed approach allows precise shaping of system response and improved regulation at the point of connection. These features make E-STATCOMs a flexible and scalable solution for integrating large data centers into weak grids and long transmission systems, supported by a design-oriented framework that facilitates parameter selection and performance assessment without extensive reliance on EMT studies to meet grid codes and AIDC interconnection requirements.

prabhat.ranjan-bana@hitachienergy.com

## 1. INTRODUCTION

The rapid growth of large-scale AI data centers (AIDC) has led to a significant increase in concentrated commercial and industrial loads in modern power systems [1], [2]. These facilities are often interconnected in regions with limited local generation and reduced short-circuit strength, leading to operational and stability challenges that require coordinated attention from both utilities and large consumers. Unlike conventional loads, data centers are interfaced to the grid through tightly controlled electrical infrastructure, including uninterruptible power supply (UPS) systems and fast-acting control loops, which can introduce complex, poorly understood dynamic interactions with the grid [3], [4].

Such interactions become particularly critical in weak network conditions, where limited system strength and reduced damping can expose latent instabilities. Field observations have shown that controller interactions within data center infrastructure can lead to oscillatory behavior, resonance phenomena, and unexpected dynamic responses, which are difficult to diagnose due to limited visibility into behind-the-meter systems [4], [5]. Moreover, the dynamic behavior of these loads is often not adequately captured in conventional planning and analysis frameworks, which typically emphasize steady-state operation and energy efficiency rather than system-level stability.

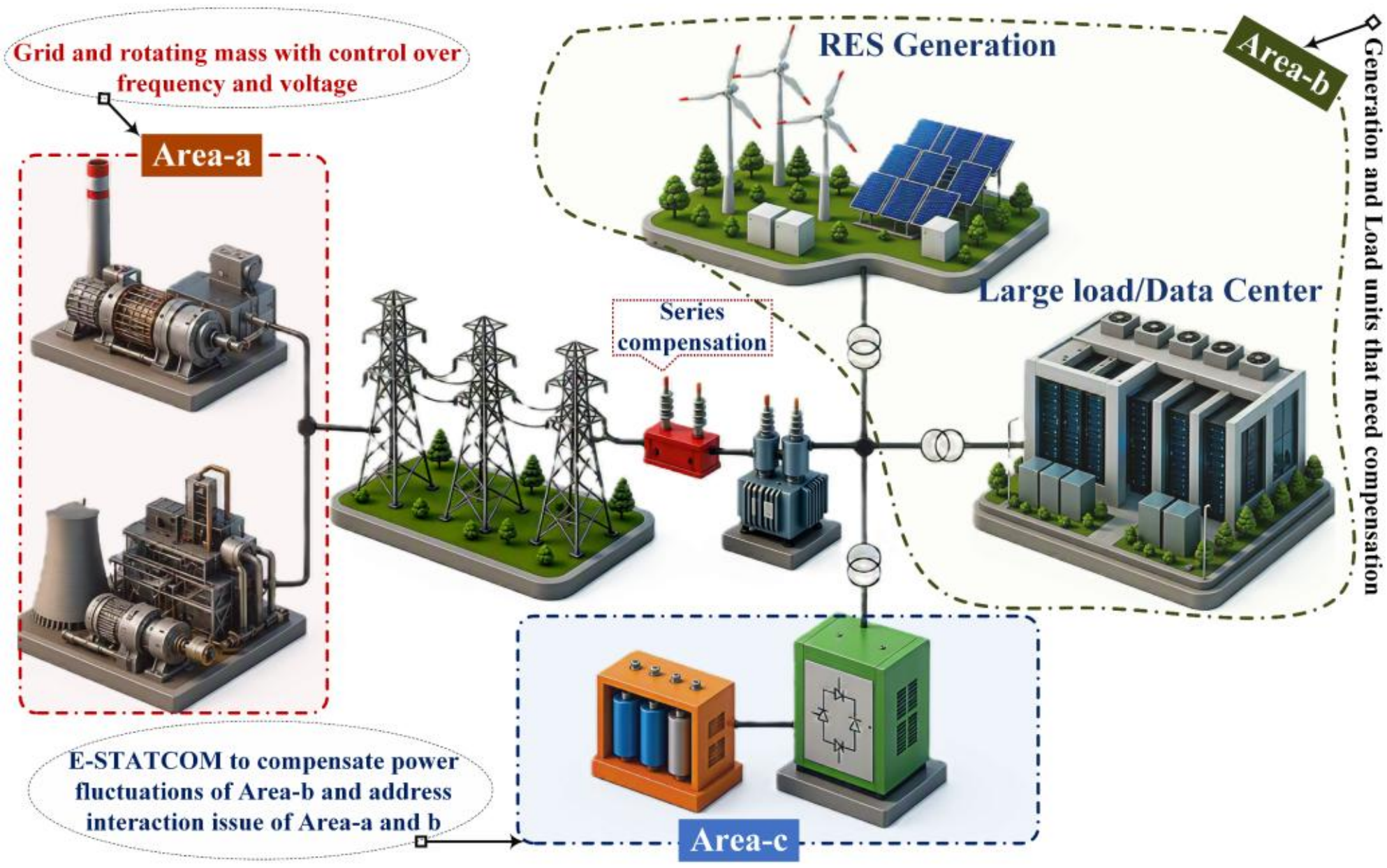

*Figure 1: Typical grid-connected system with generation, load, and compensation units*

To address these challenges, there is a growing need for grid-support solutions that can actively shape system dynamics at the point of connection [6]. In this context, grid-forming (GFM) Enhanced STATCOMs (E-STATCOMs) offer a promising approach by providing fast, controllable active power and voltage support response [7], [8]. However, their effectiveness depends strongly on the design of the grid-forming control, particularly in ensuring adequate damping and stable interaction with both the network and the data center interface.

In parallel, analyzing stability issues in such environments requires modeling approaches that capture the coupled dynamics between the grid and fast-acting load-side controls. While *dq*-domain impedance-based methods remain widely used, recent work has highlighted the advantages of alternative formulations that directly relate electrical variables to power response, enabling more intuitive interpretation of system behavior and facilitating aggregation of complex subsystems [8], [9].

Motivated by these challenges, this paper formulates a power-admittance-based linear modeling framework to study the dynamic interactions among AIDC, the grid, and compensating devices. The approach is used to analyze the small-signal behavior and quantify the impact of E-STATCOM control on system stability under varying grid conditions. The objective is to provide insight into the underlying mechanisms governing stability and to support the design of effective control strategies for reliable data center load compensation.

## 2. SYSTEM MODELLING AND ANALYSIS

The system configuration shown in Figure 1 is considered to investigate the emerging load compensation and stability challenges associated with large-scale data center integration in weak power networks and to assess suitable mitigation strategies. The setup consists of a data center connected to the grid at the point of common coupling (PCC) and supplied via a transmission corridor.

A GFM E-STATCOM is connected in shunt at the PCC to provide the required compensation. Note that the E-STATCOM is a power electronic converter device that consists of three main elements: a Modular Multilevel Converter (MMC), supercapacitor-based energy storage, and a MACH system equipped with GFM control [10]. This enables the fast active and reactive power exchange to handle the rapid demand variations typical of data center loads. The goal is to evaluate the grid-support capability of the E-STATCOM in this context, particularly its role in improving the active power profile, regulating voltage, damping, and ensuring stable operation of grid-connected data centers in weak and converter-dominated networks, while also providing insights for the systematic design and tuning of such compensating devices. In line with this, the system can be further reduced to a Thevenin equivalent, as shown in Figure 2 for simplicity.

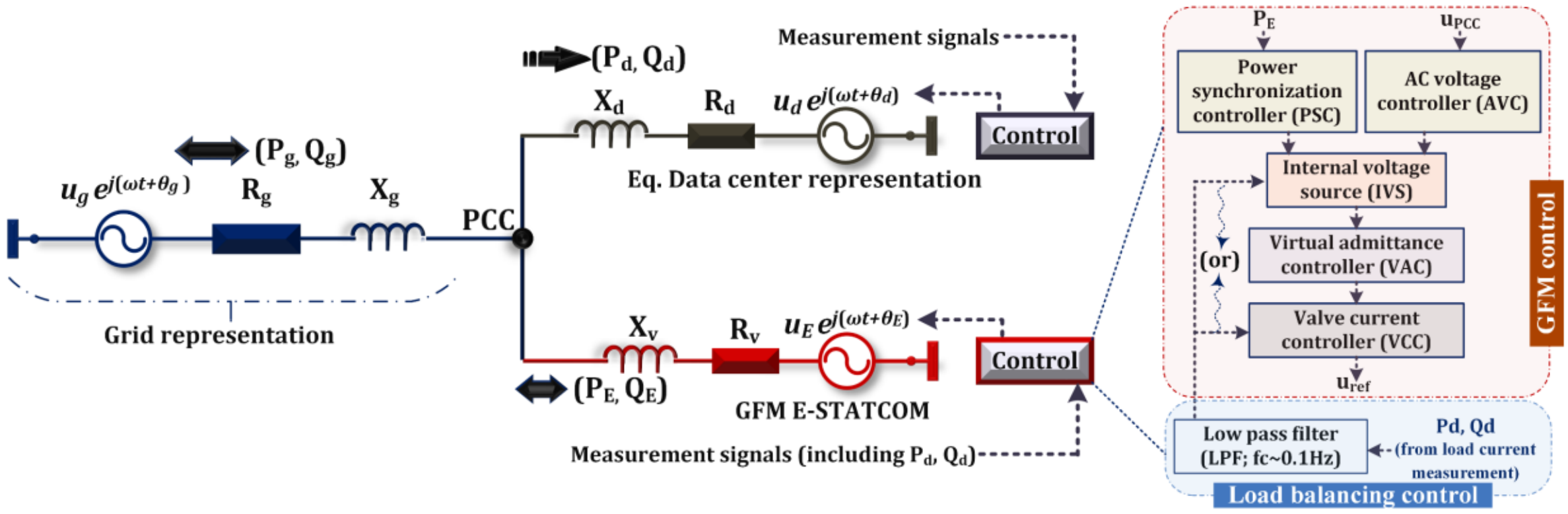

*Figure 2: Thevenin equivalent representation of a grid-connected AIDC system with E-STATCOM for linear modeling*

### 2.1 Grid-connected Data Center System

For simplicity, assume the system consists only of a data center connected to the grid in Figure 2. The grid and the data center can be represented as two voltage sources connected through their interfacing impedances. The total equivalent interfacing admittance can be written as

$$Y(s) = \frac{1}{R_t + (s + j\omega_1)L_t} = g_t(s) - jb_t(s);\; R_t = R_d + R_g \text{ and } L_t = L_d + L_g \tag{1}$$

The grid-side source voltage, the controlled data center voltage, and their angle difference can be represented as:

$$u_g = U_g e^{j\theta_g},\; u_d = U_d e^{j\theta_d} \text{ and } \delta_{gd} = \theta_g - \theta_d. \tag{2}$$

The current flowing from the grid toward the data center is:

$$i_{gd}(s) = Y_t(s)\left(u_g - u_d\right). \tag{3}$$

Accordingly, the grid active and reactive powers are:

$$P_g = U_g^2 g_t - U_g U_d\left(g_t \cos\delta_{gd} + b_t \sin\delta_{gd}\right), Q_g = -U_g^2 b_t - U_g U_d\left(g_t \sin\delta_{gd} - b_t \cos\delta_{gd}\right). \tag{4}$$

Using small-signal perturbations around an operating point, the change in grid-side power can be expressed as

$$\begin{bmatrix}\Delta P_g\\ \Delta Q_g\end{bmatrix}=\mathbf{G}_{gd}(s)\begin{bmatrix}\Delta\delta_{gd}\\ \Delta U_{gd}\end{bmatrix};\quad \begin{matrix}\Delta\delta_{gd}=\Delta\theta_g-\Delta\theta_d\\ \Delta U_{gd}=\Delta U_g-\Delta U_d\end{matrix} \tag{5}$$

Considering a particular operating point $U_{g0}\approx U_{d0}\approx U_0$, $\delta_{gd0}\approx 0$, and assuming the grid is stiff ( $\Delta\theta_g,\ \Delta U_g\approx 0$), the reduced transfer matrix becomes:

$$\begin{bmatrix}\Delta P_g\\ \Delta Q_g\end{bmatrix}=\mathbf{G}_{gd}(s)\begin{bmatrix}\Delta\theta_d\\ \Delta U_d\end{bmatrix};\ \mathbf{G}_{gd}(s)=\begin{bmatrix}-U_0^2 b_t(s) & U_0 g_t(s)\\ -U_0^2 g_t(s) & -U_0 b_t(s)\end{bmatrix};\text{ where}$$

$$g_t(s)=\frac{R_g+R_d+s(L_g+L_d)}{\left(R_g+R_d+s(L_g+L_d)\right)^2+\omega_1^2(L_g+L_d)^2};\ b_t(s)=\frac{\omega_1(L_g+L_d)}{\left(R_g+R_d+s(L_g+L_d)\right)^2+\omega_1^2(L_g+L_d)^2}. \tag{6}$$

Note that the input matrix of (6) is obtained through a power and voltage controller in reality. Assuming a standard 2nd order power controller and a 1st order voltage controller, the control inputs can be expressed as:

$$\begin{aligned}&\Delta\theta_d(s)=C_{P,d}(s)\left(\Delta P_{\text{ref},d}(s)-\Delta P_d(s)\right);\ C_{P,d}(s)=\left(K_{p,d}+\frac{K_{i,d}}{s}\right)\frac{1}{s}\\ &\Delta U_d(s)=C_{V,d}(s)\Delta U_d^{\star}(s);\ C_{V,d}(s)=\left(K_{pV,d}+\frac{K_{iV,d}}{s}\right)\end{aligned} \tag{7}$$

Substituting (7) into (6), the final transfer function can be expressed as:

$$\begin{bmatrix}\Delta P_g\\ \Delta Q_g\end{bmatrix}=\begin{bmatrix}\dfrac{-U_0^2 b_t(s)C_{P,d}(s)}{1+U_0^2 b_t(s)C_{P,d}(s)} & \dfrac{U_0 g_t(s)C_{V,d}(s)}{1+U_0^2 b_t(s)C_{P,d}(s)}\\ \dfrac{-U_0^2 g_t(s)C_{P,d}(s)}{1+U_0^2 b_t(s)C_{P,d}(s)} & \dfrac{-U_0 b_t(s)-U_0^3\left(g_t^2(s)+b_t^2(s)\right)C_{P,d}(s)}{1+U_0^2 b_t(s)C_{P,d}(s)}C_{V,d}(s)\end{bmatrix}\begin{bmatrix}\Delta P_{\text{ref},d}\\ \Delta U_d^{\star}\end{bmatrix} \tag{8}$$

The above admittance matrix enables frequency-domain analysis at a specific operating point, as it is obtained by linearizing the system around a fixed equilibrium point [9]. However, the system's resonance characteristics are sensitive to the initial operating conditions. Variations in initial conditions (*i.e.*, if $U_{g0}\neq U_{d0}\neq U_0$ or $\delta_{gd0}\neq 0$) can shift the resonance frequency, thereby limiting the accuracy of (8). To capture this dependency, initial operating conditions are incorporated into (8), yielding a modified linearized admittance matrix expressed as:

$$\begin{bmatrix}\Delta P_g\\ \Delta Q_g\end{bmatrix}=\begin{bmatrix}\dfrac{a_{p\theta}(s)C_{P,d}(s)}{1-a_{p\theta}(s)C_{P,d}(s)} & \dfrac{a_{pU}(s)C_{V,d}(s)}{1-a_{p\theta}(s)C_{P,d}(s)}\\ \dfrac{a_{q\theta}(s)C_{P,d}(s)}{1-a_{p\theta}(s)C_{P,d}(s)} & \dfrac{a_{qU}(s)\left(1-a_{p\theta}(s)C_{P,d}(s)\right)+a_{q\theta}(s)C_{P,d}(s)a_{pU}(s)}{1-a_{p\theta}(s)C_{P,d}(s)}C_{V,d}(s)\end{bmatrix}\begin{bmatrix}\Delta P_{\text{ref},d}\\ \Delta U_d^{\star}\end{bmatrix} \tag{9}$$

where,

$$\begin{aligned}&a_{p\theta}(s)=U_{g0}U_{d0}\left(g_t(s)\sin\delta_{gd0}-b_t(s)\cos\delta_{gd0}\right),\ a_{pU}(s)=U_{g0}\left(g_t(s)\cos\delta_{gd0}+b_t(s)\sin\delta_{gd0}\right),\\ &a_{q\theta}(s)=-U_{g0}U_{d0}\left(g_t(s)\cos\delta_{gd0}+b_t(s)\sin\delta_{gd0}\right),\ a_{qU}(s)=U_{g0}\left(g_t(s)\sin\delta_{gd0}-b_t(s)\cos\delta_{gd0}\right).\end{aligned}$$

## 2.2 Frequency and Time-Domain Analysis

The analysis in this subsection is based on a multi-operating linearized model expressed in (9). Note that the perturbation or disturbance is applied to the data center active power reference ($P_{ref,d}$). Typically, conventional power-admittance formulations are expressed in terms of the active-power response to frequency or phase-angle perturbations [8], [9], which is appropriate for analyzing generation units and frequency-support mechanisms. In contrast, large data centers behave primarily as controllable loads, with a dominant disturbance in power demand. Therefore, this work focuses on the transfer characteristics between data-center power variations and the resulting grid power response. This representation provides a direct assessment of how load fluctuations propagate through the network, how resonant modes shift with operating conditions, and how the E-STATCOM influences these interactions.

In line with this, Figure 3(a) shows the grid power ($P_g$) response corresponding to the perturbations in the reference power of AIDC ($P_{ref,d}$). The results clearly show that the $P_g$ experiences sub-synchronous resonance (SSR) and DC, whose dynamics depend on the initial operating point. In particular, changes in the system's initial power flow directly shift the resonance frequency. As the initial loading increases, the dominant resonance moves to lower frequencies. This shift is consistently observed as sub-synchronous oscillation (SSO) in the corresponding time-domain response shown in Figure 3(b) with a 0.5 p.u. of step change in $P_{ref,d}$. Note that, depending on the power controller design, the SSR may appear in the range of 3~30Hz, which is well within the damping range of the E-STATCOM.

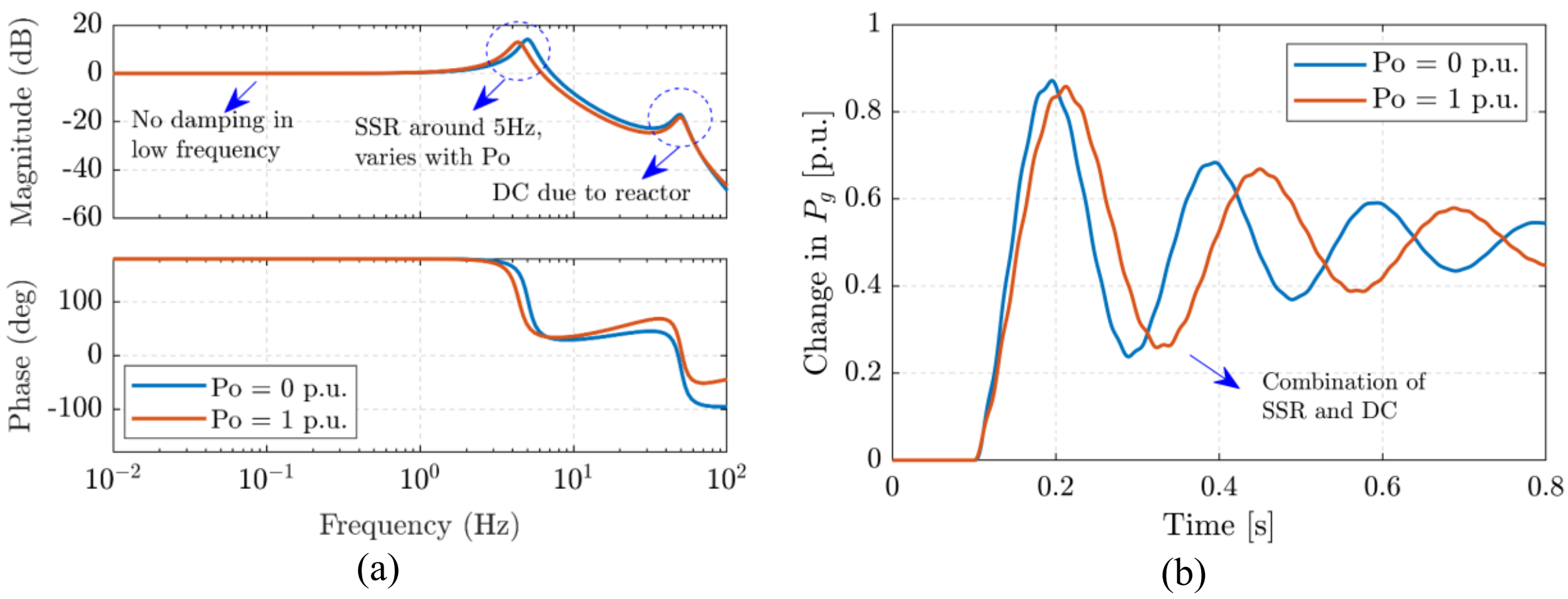

*Figure 3: Power response of the grid-connected AIDC system using expression (9), (a) Frequency-domain results of change in $P_g$ with perturbations in $P_{ref,d}$, (b) Time-domain response of $P_g$ with 0.5 p.u. change in $P_{ref,d}$*

These observations indicate that the system cannot be characterized by a single operating point. The dependence of resonance behavior on data center loading must be explicitly considered when analyzing stability and designing grid-support solutions.

## 3. COMPENSATION WITH E-STATCOM

GFM E-STATCOMs have been shown to be a flexible means of supplying inertia and supporting the integration of renewables [9], [11]. By combining fast converter control with an energy storage interface, the E-STATCOM can actively regulate voltage and power exchange at the point of connection, thereby improving overall system stability under rapid load variations. The following subsections discuss the integration and performance of the E-STATCOM for AIDC using the linear modeling framework described above.

### 3.1 Integration of E-STATCOM

The E-STATCOM can be integrated into the grid-connected AIDC system using the formulation in (9). This can be achieved by considering the PCC as the reference point for the individual subsystems (Grid, AIDC, and E-STATCOM). In this work, the E-STATCOM is represented as a controlled voltage source behind an equivalent impedance, consistent with the behavior expected from a grid-forming converter [12]. In addition, a load-balancing function similar to that used for arc furnace applications for flicker control is incorporated into the GFM control of the E-STATCOM to provide load smoothing [13]. The schematic of the overall E-STATCOM control is represented in Figure 2.

In line with the above control structure, the virtual impedance parameters are selected as $R_v = X_v = 0.2$ p.u. The power control loop is configured with an inertia constant of H = 10 s and a damping factor of 0.7. The voltage controller is operated slowly. Furthermore, the load-balancing function is designed with a cutoff frequency of 0.01 Hz, implying that load fluctuations above this frequency are compensated by the E-STATCOM, while slower variations are supplied directly by the grid. Choosing a further lower cutoff frequency would require the E-STATCOM to exchange energy over longer durations, resulting in a larger energy storage requirement.

It should be noted that the power and energy rating of an E-STATCOM is a project-specific techno-economic optimization problem that depends on the load profile, compensation objectives, smoothing bandwidth, and energy storage technology. In this study, the E-STATCOM is assumed to have an energy rating equal to the rated data-center load. The focus of the paper is therefore on dynamic behavior and controller design rather than optimal sizing.

### 3.2 Comparative Frequency and Time-Domain Analysis

Figure 4(a) presents the frequency-domain response of the grid active power $P_g$ to variations in the AIDC power reference $P_{ref,d}$. The response can be divided into four distinct regions based on the dominant E-STATCOM control action. In region-a, corresponding to frequencies below the load-balancing cutoff frequency of 0.01 Hz, the E-STATCOM does not actively compensate power fluctuations. Therefore, the magnitude response of $P_g$ remains essentially identical with and without the E-STATCOM. These low-frequency variations must be supplied by the grid itself or by other nearby assets, such as BESS or other generation sources. Starting from region-b, the load-balancing controller becomes active and the E-STATCOM begins compensating load fluctuations. As a result, the converter exhibits a synchronizing or stiffness-type response, reducing the magnitude of disturbances propagating to the upstream grid.

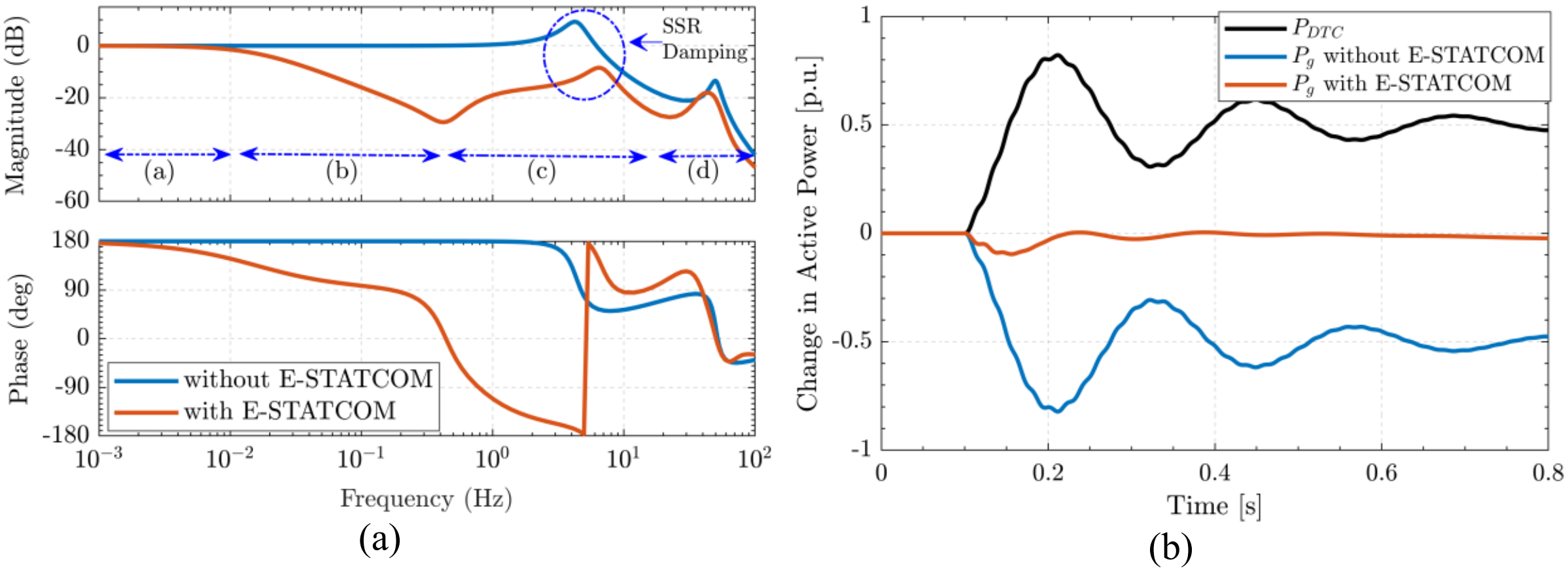

*Figure 4: Power response of the grid-connected AIDC system with E-STATCOM, (a) Response of $P_g$ with perturbations in $P_{ref,d}$, (b) Time-domain response of $P_g$ with 0.5 p.u. change in $P_{ref,d}$*

In region-c, the power synchronization loop becomes dominant. Consequently, the inertia emulation (H=10 s) of E-STATCOM produces a damping response to the system dynamics. This effect becomes particularly important around the SSR frequency, where oscillatory behavior is suppressed. Finally, region-d is increasingly influenced by the virtual admittance (VAC). Although the VAC remains active across the entire frequency range, its impact becomes more pronounced in this region, where it shapes the converter impedance and contributes to the overall damping characteristics.

The combined action of these control functions significantly modifies the system response. As shown in Figure 4(a), the uncompensated system exhibits a pronounced SSR mode around 5 Hz. With the E-STATCOM connected at the PCC, this mode is substantially damped, indicating improved dynamic stability and reduced sensitivity to load-induced disturbances. The effect of this damping is further confirmed by the time-domain response shown in Figure 4(b). Following a step increase of 0.5 p.u. in $P_{ref,d}$, the oscillations in $P_g$ are considerably reduced, resulting in a faster and better-damped response.

## 4. SOFTWARE-IN-THE-LOOP (SIL) WITH DETAILED EMT MODEL

SIL analysis is conducted within the EMT environment to validate the theoretical concepts and mathematical models previously discussed. The system under study incorporates a detailed network featuring a realistic grid, an AIDC load model, and the detailed E-STATCOM model. The AIDC load model is developed based on the generic EMT AIDC model library in PSCAD, published by PNNL [14]. The model represents the following components:

- IT-related equipment: Represented with a detailed load curve, which has ramp-up and ramp-down rates, dynamic power range, and critical frequency range.
- Power delivery systems:
  - Two transformers: high to medium voltage and Medium to Low voltage
  - UPS system for IT equipment: it is a centralized double conversion UPS. Both the UPS rectifier and inverter are voltage-source converters, represented by an average-value model (AVM). The input rectifier control is grid-following (GFL) with current vector control. The output inverter control is GFM with droop control. The battery converter is a bidirectional buck-boost converter. The modulation index of the converter is controlled using both an active power regulator (to manage charging and discharging) and a DC voltage regulator
- Cooling systems:
  - Chiller pump load; it is modeled as a power electronic load, including variable speed drives and inverters
  - Fan load; it is modeled as a power electronic load, including variable speed drives and inverters
- Miscellaneous is represented with constant impedance loads.

### 4.1 Sub-synchronous Oscillation (SSO) Analysis

To evaluate the SSO damping capability of the E-STATCOM, the load power controller is tuned to a bandwidth of approximately 5 Hz, while the load disturbance occurs at 0.25 Hz. Figure 5 depicts that a variation in load power excites a 5 Hz oscillatory mode. Without compensation, this oscillation appears directly in the grid power at the point of interconnection (POI), as shown in the 2nd subplot. With the E-STATCOM connected, the oscillatory component is largely absorbed by the compensator, as illustrated in the 3rd subplot. As a result, the power seen by the grid remains smooth, and the SSO is damped out (see 2nd subplot), confirming that the E-STATCOM effectively damps the SSO and prevents load-induced oscillations from propagating upstream.

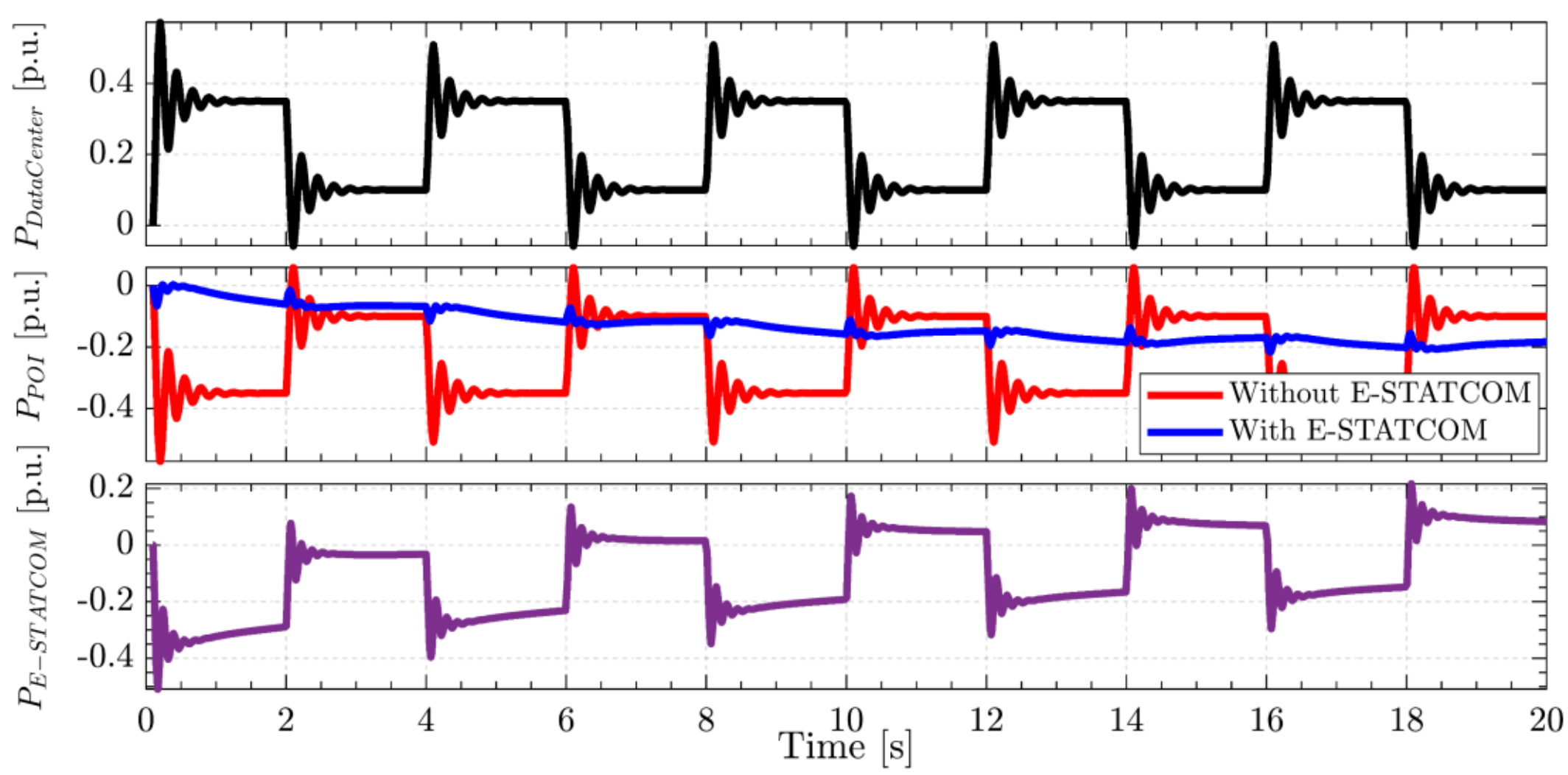

*Figure 5: Data center active power, Grid active power/power at POI and E-STATCOM output, for a 0.25Hz varying load with SSO of 5Hz*

### 4.2 Fault Ride-through

A three-phase fault is applied at the data center POI, producing a deep voltage sag that exceeds the undervoltage pickup threshold and causes the AIDC to disconnect from the grid and transfer its load to local backup generation. From the grid perspective, this sudden disconnection appears as a large load loss, which can be problematic depending on network strength and may lead to over-frequency and cascading outages. A recent NERC incident report [15] documented this risk, reporting the simultaneous loss of approximately 1,500 MW of voltage-sensitive load following a transmission fault, which caused a rise in system frequency and required operator action to control voltage excursions. These risks motivate utilities and TSOs to define ride-through requirements for large loads, including data centers, requiring them to remain connected as long as the POI voltage stays within the applicable voltage ride-through envelope. For example, the ATC criteria in [16] specify that during a deep voltage sag approaching zero voltage, the load must remain online and may trip only if the sag duration exceeds 150 ms.

In accordance with the above requirement, Figure 6 shows that, without compensation, the data center load trips and remains offline after the fault. With the E STATCOM in service, the system recovers rapidly after fault clearing and restores the POI load close to its pre-fault level. This enables E-STATCOM to comply with ride-through requirements during the simulated event.

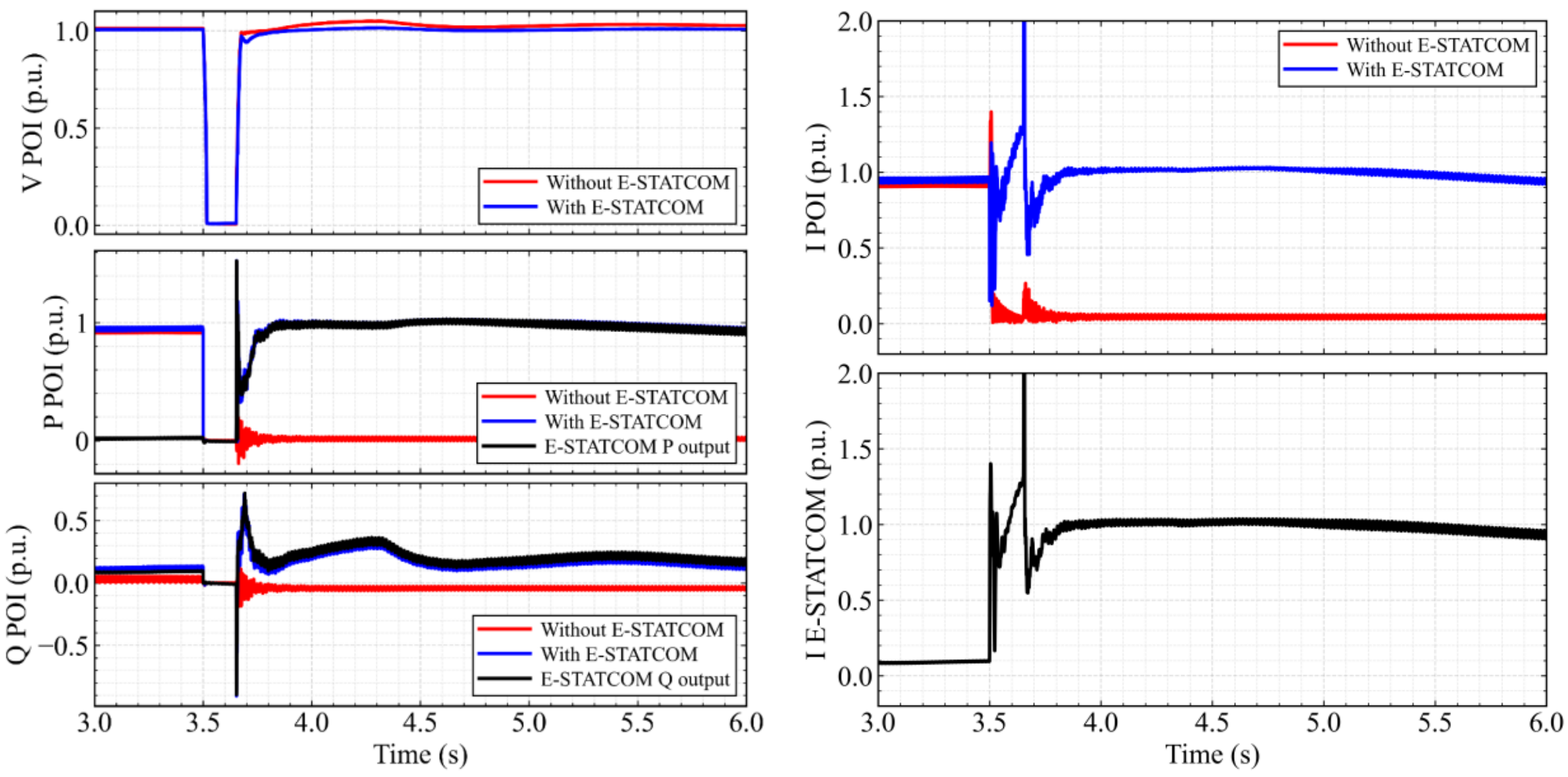

Figure 6: Fault ride-through performance of the AIDC system following a 3-phase fault at POI with and without E-STATCOM

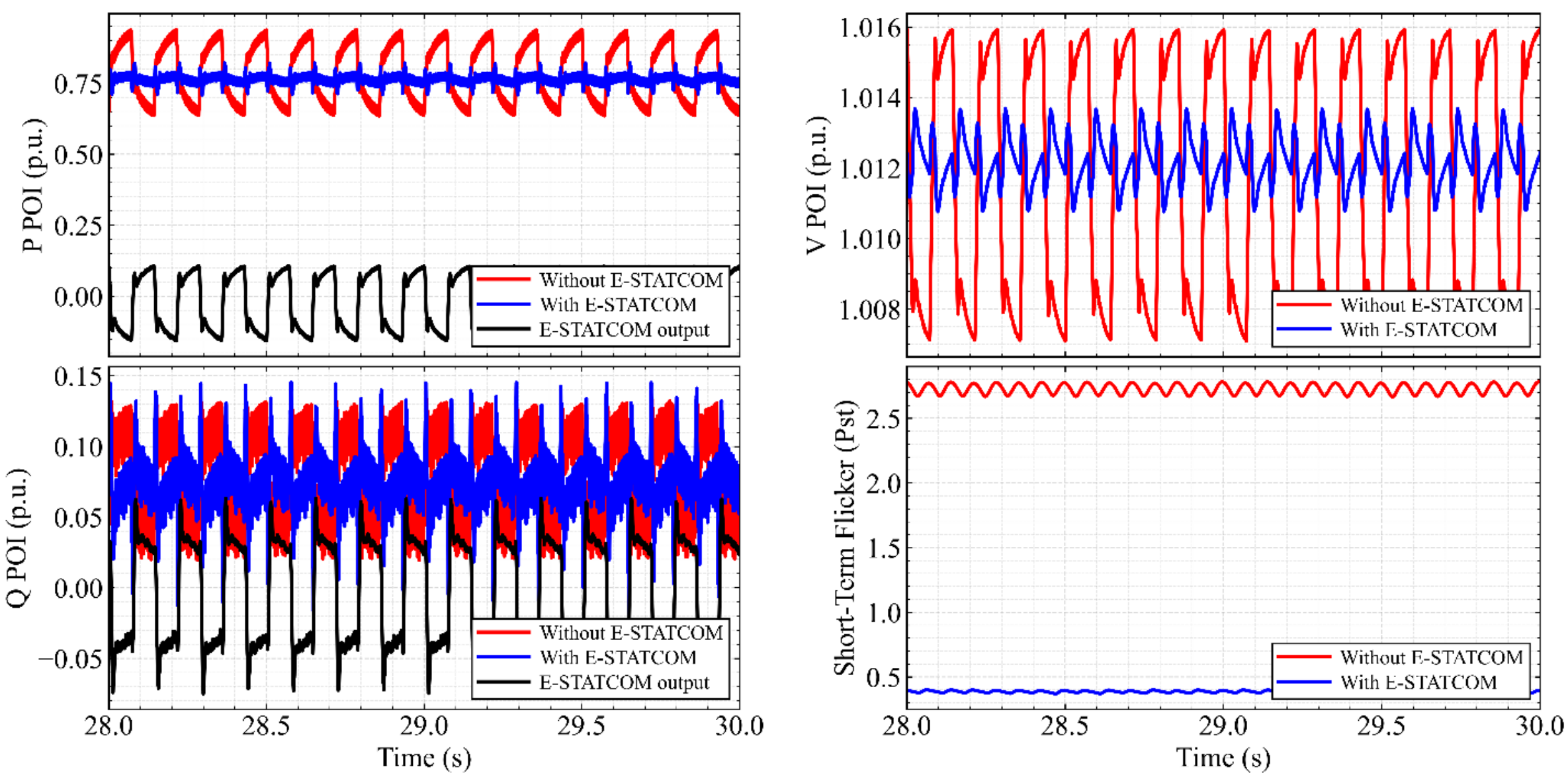

Figure 7: Active & reactive power, voltage, and short-term flicker measured at data center POI, with and without E-STATCOM

### 4.3 Flicker Analysis

The E-STATCOM's ability to mitigate flicker is assessed for an AIDC load exhibiting approximately 40% variable load oscillating at 7 Hz. Without compensation, the resulting active power fluctuations cause voltage variations at the POI and increase the simulated short-term flicker level (Pst) to about 2.6, exceeding the IEEE 1453 planning limit of 0.8 [17]. With the E-STATCOM in service, fast load fluctuations are effectively suppressed, and the Pst level remains within the prescribed limit of 0.8, as illustrated in Figure 7

### 4.4 Analysis with Realistic Load Profile

To assess GFM E-STATCOM performance under a more realistic AIDC load with multiple frequency components, an EMT simulation is performed using the NERC load profile [2]. Figure 8(a) shows the POI active power for both uncompensated and compensated cases. It can be observed that the GFM E-STATCOM effectively smooths the load response, preventing large oscillations from propagating to the POI.

Further, the FFT profile in Figure 8(b) confirms that multiple frequency components are well damped, demonstrating the E-STATCOM's ability to mitigate a broad range of fluctuation-driven risks, including SSO, local and inter-area oscillations, and power quality issues. What's more, the AIDC with NERC profile produces significant Pst variations over time, as shown in Figure 8(c). The E-STATCOM compensates the flicker to a desirable level, as specified by the IEEE 1453 standard.

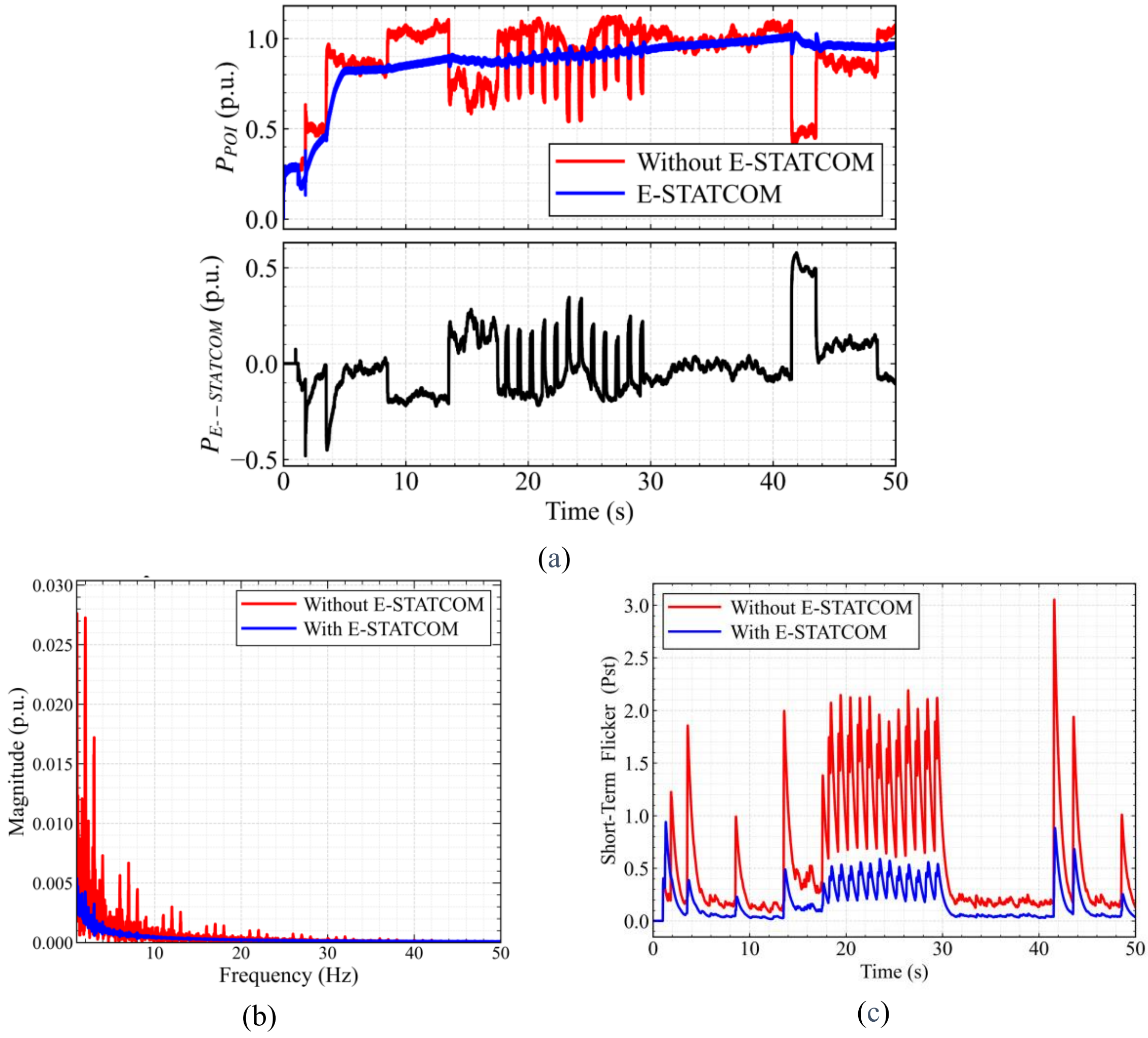

*Figure 8: Performance with NERC profile (a) Active power at data center POI, E-STATCOM output, (b) FFT of the active power, with and without E-STATCOM, (c) Pst profile with and without E-STATCOM*

## 4.5 Impact of SSO on Nearby Generator Shaft

AIDC load fluctuations can excite SSO between the electrical network and turbine-generator mechanical modes, typically in the 5 Hz to over 30 Hz range [2], [3]. Disturbances near these natural frequencies, including generation change, load oscillations, or forced periodic load variations, can induce high-amplitude rotor shaft oscillations and mechanical stress. Unlike short-duration conventional transients, sustained or repetitive AI load fluctuations may accumulate fatigue damage over time, increasing the risk of shaft fatigue or failure even when individual torque cycles are relatively small.

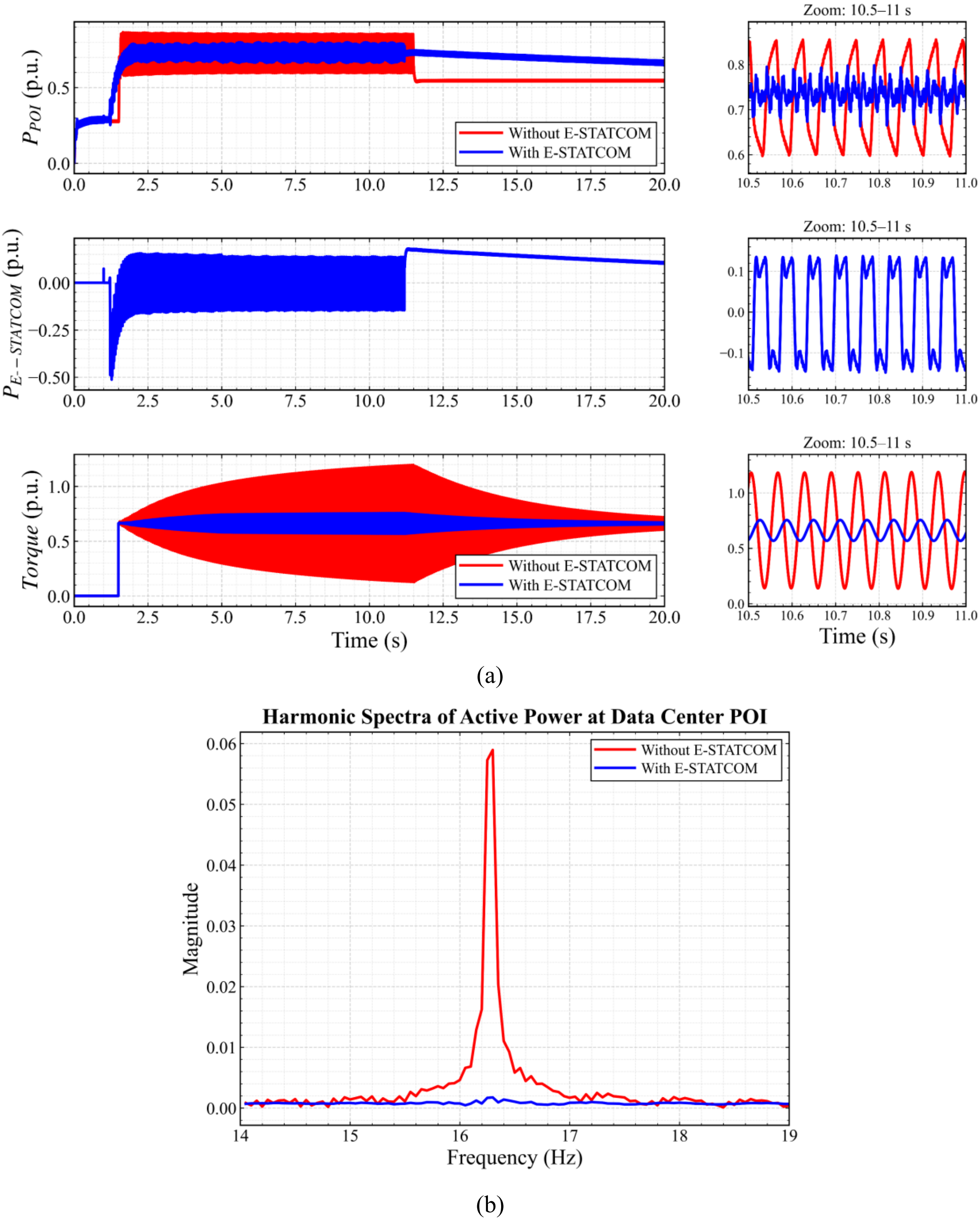

*Figure 9: Simulation results of (a) Data center active power, E-STATCOM output, and a synchronous generator shaft torque, with and without E-STATCOM, and (b) FFT of data center active power with and without E-STATCOM*

Further to the above, the ability of the GFM E-STATCOM to suppress AI load oscillations and reduce the risk of SSO in a nearby synchronous generator is evaluated. The data center load alternates between idle and training cycles, resulting in power oscillations between 0.6 p.u. and 1.0 p.u. at 16 Hz. This frequency corresponds to one of the generator's torsional modes. As presented in Figure 9, the E-STATCOM effectively damps these oscillations, substantially reducing generator shaft torque oscillations compared with the uncompensated case.

## 5. CONCLUSION

This paper presents a power-admittance-based linear modeling framework for analyzing the dynamic interactions among large-scale data centers, the grid, and grid-forming E-STATCOM. Unlike conventional formulations based on phase-angle perturbations, the developed approach evaluates the grid response directly with respect to variations in data center power demand, thereby providing a more relevant representation of large-load behavior. The analysis showed that the system's resonance characteristics are strongly dependent on the operating point and can shift with changes in data center loading.

Based on the developed framework, a grid-forming E-STATCOM with load-balancing functionality was integrated into the system and evaluated in both frequency and time domains. The results demonstrated that the E-STATCOM effectively compensates load fluctuations, reduces the propagation of disturbances to the upstream grid, and provides substantial damping of the dominant SSR oscillatory mode.

The theoretical findings were further validated through detailed EMT simulations that incorporated a realistic data center model and representative AI load profiles. The E-STATCOM was shown to smooth active power variations, suppress oscillatory components over a wide frequency range, mitigate flicker to an acceptable range and reduce the risk of SSOs in nearby synchronous generators. Overall, the study demonstrates that grid-forming E-STATCOM can provide an effective solution for supporting the integration of large-scale data centers into weak power systems while maintaining stable and reliable grid operation.